\documentstyle[12pt]{article}
\renewcommand{\theequation}{\thesection.\arabic{equation}}
\begin{document}
\title{ Justification of the zeta function \\renormalization in rigid
string model} \author{ V.V.~Nesterenko and
I.G.~Pirozhenko\thanks{Permanent address:
Petrozavodsk State University, Petrozavodsk, 185640, Russia}\\
{\small \it Bogoliubov Laboratory of Theoretical Physics}\\
 {\small \it Joint Institute for Nuclear Research,
Dubna, 141980, Russia.}}
\date{}
\maketitle
\begin{abstract}
A consistent procedure for regularization of divergences and for the
subsequent renormalization of the string tension is proposed in
the framework of the one-loop calculation of the interquark
potential generated by the
Polyakov-Kleinert string. In this way, a justification of the formal
treatment of  divergences by analytic continuation of the Riemann
and Epstein-Hurwitz zeta functions is given. A spectral representation
for the renormalized string energy at zero temperature is derived,
which enables one to find the Casimir energy in this string model
at nonzero temperature very easy.
\end{abstract}
\newpage
\section{Introduction}
A consistent method to treat the divergences in quantum field theory
is known to be the following~[1]. Divergent expressions must
at first be regularized, for example, by the Pauli-Willars method,
then subtractions justified by transition to physical (observable)
parameters of the theory should be done. After that the regularization
is to be removed.

Together with this approach there are widely used methods that do not
apply explicit regularization and renormalization but which
nevertheless give finite answer. First of all, it is the zeta
function technique. The main idea of this approach  is the
following~[2-4]. One assumes that the divergent
sum $\sum_n \omega_n$ of eigenvalues of the operator\footnote{Most
commonly this operator is the Laplace operator $(-\triangle)$. The
zeta function regularization is usually applied to the Euclidean
version  of the models where one has to do with elliptic
operators~[3].}
determining the dynamics in the model under consideration
is equal to the value of the zeta function
for this operator, $\zeta(s)$, when $s\to-1$. At first the function
$\zeta(s)$ is defined by the formula $\zeta(s)=\sum_n \omega_n^{-s}$
for $\mbox{Re}\,s>1$, and then it is analytically
continued to $\mbox{Re}\,s\le 1$
possibly save for isolated points. In the case of the Dirichlet
boundary conditions for the two-dimensional Laplace and Helmholtz
operators this function appears to be the Riemann
$\zeta$-function or the Epstein-Hurwitz $\zeta$-function,
respectively.
These functions are widely used in calculations of the Casimir energy
in field~[4,5] and string models~[6].

Undoubtedly such a formal method to treat divergences needs
justification in each particular case~[7,8]. The more so, there are
examples when analytic continuation leads to ambiguities~[9]. To
justify this approach, it is necessary to show that it gives the same
results as the standard renormalization procedure with regularization
and subtraction.
It is this problem that  will be considered in the present paper in
the framework  of one-loop calculation of  the interquark potential
(or the Casimir energy) in the rigid string model. This model is
chosen
because here both the Riemann and Epstein-Hurwitz $\zeta$-functions
are employed.

The interquark potential generated by a rigid string was studied in a
number of papers by making use of the perturbation theory and
variational estimation of the functional integral (see, for example,
Ref.~[10] and papers cited therein). These results are well-known.
Therefore attention will be basically  paid to development of the
consistent procedure of renormalization and to justification, on
this basis, the results obtained by $\zeta$-function method.

The layout of the paper is as follows. In Section 2, the interquark
potential generated by a rigid string is calculated in the one-loop
approximation, the standard method of analytic continuation of the
Riemann and Epstein-Hurwitz $\zeta$-functions being used.
In Section 3,
the consistent regularization of the divergences and the string
tension renormalization are carried out. Unlike the $\zeta$-function
method, the finite expression for the string potential is derived
here uniquely. Moreover in our approach the renormalized string
energy at zero temperature is obtained in terms of the spectral
representation that can be directly generalized to a finite
temperature. In the Conclusion (Section 4), the obtained results are
discussed in short. Auxiliary material concerning the details of
the calculations is given in Appendices A and B.

\section{Interquark potential generated by rigid string in
one-loop approximation} \setcounter{equation}{0}

We consider the most simple example of the application of the
Riemann and Epstein-Hurwitz $\zeta$-functions.
This is the calculation of the interquark
potential generated by the Polyakov-Kleinert string [11,12] in
the one-loop approximation~[13].
In spite of its simplicity, this example demonstrates the main
features  of the approach.

For our purpose the quadratic approximation to the Polyakov-Kleinert
string action is sufficient
\begin{equation}
S^{\beta}=M_0^2\int\limits_{0}^{\beta}dt\int\limits_{0}^{R}dr\left[1+
\frac{1}{2}{\bf u}\left(1-\frac{\alpha}{M_0^2}\triangle\right)
\left(-\triangle\right){\bf u}\right].
\label{a2.1}
\end{equation}
Here $M_0^2$ is the string tension,
${\bf u}(t,r)=(u^1(t,r),u^2(t,r),\dots,u^{D-2}(t,r))$ are the
transverse string coordinates in \mbox{$D$-dimensional} space-time,
and
$\alpha$ is a dimensionless parameter characterizing the string
rigidity, $\alpha>0$, $R$ is the distance between quarks connected by
string, i.e.,\ the string length.
The Euclidean action is considered, therefore the operator
$\triangle$ in (\ref{a2.1}) is the two-dimensional Laplace operator
$\triangle=\partial^2/ \partial t^2+\partial^2/\partial r^2$.  The
"time" variable $t$ ranges in the interval $0\le t \le\beta$,
where $\beta=1/T$ is the inverse temperature.

The action (\ref{a2.1}) should be completed by boundary conditions for
string coordinates at the points $r=0$ and $r=R$. Usually a string
with fixed ends is considered
\begin{equation}
{\bf u}(t,0)={\bf u}(t,R)=0\,{.}
\label{a2.2}
\end{equation}
This corresponds to the static interquark potential.
The string potential $V(R)$ is defined in a standard way
\begin{equation}
\exp[-\beta V(R)]=\int[D{\bf u}]\exp{\left\{-S^{\beta}[{\bf u}]
\right\}},
\;\;\beta\to\infty.
\label{a2.3}
\end{equation}
The functional integral in (\ref{a2.3}) is taken over string
coordinates ${\bf u}(t,r)$ that satisfy periodic conditions in
the time  variable $t$
\begin{equation}
{\bf u}(t,r)={\bf u}(t+\beta,r).
\label{a2.4}
\end{equation}

Inserting (\ref{a2.1}) into (\ref{a2.3})  one
obtains after the functional integration  when $\beta\to\infty$
\begin{equation}
V(R)=M_0^2
R+\frac{D-2}{2 \beta}\mbox{Tr} \ln(-\triangle)+ \frac{D-2}{2 \beta}
\mbox{Tr}
\ln\left(1-\frac{\alpha}{M_0^2}\triangle\right).
\label{a2.5}
\end{equation}
For calculating the traces in (\ref{a2.5}) the eigenvalues of the
operators $(-\triangle)$  and $[1-(\alpha/M_0^2) \triangle]$ with the
boundary conditions (\ref{a2.2}) and the periodicity conditions
(\ref{a2.4})
are needed
$$(-\triangle)\phi_{nm}=\lambda_{nm}
\phi_{nm},$$ \begin{equation}
\left(1-\frac{\alpha}{M_0^2}\triangle\right)\psi_{kl}=\xi_{kl}
\psi_{kl},
\label{a2.6}
\end{equation}
Using the Fourier expansion we find~[13]
$$\lambda_{nm}=\Omega_n^2+\omega_m^2,$$
\begin{equation}
\xi_{nk}=\Omega_n^2+\tilde\omega_k^2,
\label{a2.7}
\end{equation}
where $\Omega_n=2\pi n/\beta,\;\;n=0,\pm 1,\pm 2,\dots$ are the
Matsubara  frequencies, $\omega_m=m\pi/R,\;m=1,2,\dots$ are
positive roots of the
equation
\begin{equation}
\sin(\omega R)=0,
\label{a2.8}
\end{equation}
and $\tilde\omega_k=\sqrt{(k\pi/R)^2+M_0^2/\alpha},\;\;k=1,2,\dots $
are those of the equation
\begin{equation}
\sin\left(R\sqrt{\tilde\omega^2-M_0^2/\alpha}\right)=0.
\label{a2.9}
\end{equation}
Summation over the Matsubara frequencies $\Omega_n$ can be
accomplished
by making use of the known methods~[13]. Upon taking the limit
$\beta\to\infty$, the potential generated by the string assumes the
form
\begin{equation}
V(R)=M_0^2 R+(D-2)(E_C^{(1)}+E_C^{(2)}),
\label{a2.10}
\end{equation}
where $E_C^{(1)}$ ¨ $E_C^{(2)}$ are the Casimir energies corresponding
to both the modes of the rigid string oscillations
\begin{equation}
E_C^{(1)}=\frac{\pi}{2 R}\sum\limits_{n=1}^{\infty}n,
\label{a2.11}
\end{equation}
\begin{equation}
E_C^{(2)}=\frac{\pi}{2 R}\sum\limits_{n=1}^{\infty}\sqrt{n^2+a^2},~~
a^2=\frac{M_0^2 R^2}{\alpha \pi^2}.
\label{a2.12}
\end{equation}
Summation of the divergent series (\ref{a2.11}) and (\ref{a2.12})
by  analytic continuation of the $\zeta$-function  is now commonly
used.
Nevertheless we remind the main steps of this approach in short.

We begin with the first sum (\ref{a2.11}). According to the scheme
outlined in the Introduction, we  first have to consider the function
\begin{equation}
\zeta(s)=\sum\limits_{n=1}^{\infty}n^{-s},\;\;\mbox{Re}~s>1,
\label{a2.13}
\end{equation}
and then to continue it analytically to the region $\mbox{Re}~s<1$.
In this case $\zeta(s)$ is the Riemann $\zeta$-function.
Analytic continuation of
the function (\ref{a2.13}) to the rest of the complex plane $s$,
with the exception of the point $s=1$, is performed by the
contour integral~[14]
\begin{equation}
\zeta(s)=-\frac{\Gamma(1-s)}{2 \pi i}\int\limits_{C}{}
\frac{(-z)^{s-1}}{1-e^{-z}}dz,
\label{a2.14}
\end{equation}
where the contour $C$ is shown in Fig.~1. This contour should avoid
the  points $z=\pm 2 n \pi i\;\;(n=1,2,3,\dots)$. Because of
the multiplier
$\Gamma(1-s)$ in (\ref{a2.14})
the Riemann $\zeta$-function has a simple pole at $s=1$ with the
residue equal to $1$
\begin{equation}
\zeta(s)=\frac{1}{s-1}+\gamma+\dots,
\label{a2.15}
\end{equation}
where $\gamma$ is the Euler constant
\begin{equation}
\gamma=\lim\limits_{N\to\infty}\left(\sum\limits_{n=1}^{N}\frac{1}{n}-
\ln N
\right).
\label{a2.16}
\end{equation}
The $\zeta$-function defined by integral (\ref{a2.14}) satisfies the
reflection formula~[14]
\begin{equation}
\zeta(1-s)=2~(2\pi)^{-s}\cos\left(\frac{\pi s}{2}\right)\Gamma(s)
\zeta(s).
\label{a2.17}
\end{equation}

According to the scheme outlined above we have to attribute
the value $\zeta(-1)$ to the sum
 of the divergent series (\ref{a2.11}) . For $s=-1$
the integral representation (\ref{a2.14}) gives
\begin{equation}
\zeta(-1)=\frac{1}{2\pi i}\int\limits_{C}^{}\frac{(-z)dz}{z^3
\left(1-e^{-z}\right)}.
\label{a2.18}
\end{equation}
Since the integrand is single-valued in the plane
$z$, the integration contour  in Fig.~1 can be closed. As a result,
$\zeta(-1)$ is equal to the integrand residue at $z=0$. To find the
residue,
we can use the definition of Bernoulli numbers~[14]
$$ \frac{t}{e^t-1}=1-\frac{1}{2}t+B_1\frac{t^2}{2}-B_2\frac{t^4}{4!}+
\dots,$$
where $B_1=1/6,~B_2=1/30,\dots$. Thus we obtain
\begin{equation}
\zeta(-1)=-B_1\frac{1}{2}=-\frac{1}{12}.
\label{a2.19}
\end{equation}
Finally we attribute  the value
\begin{equation}
E_C^{(1)}=\frac{\pi}{2 R}\sum\limits_{n=1}^{\infty}n=
\frac{\pi}{2 R}\zeta(-1)=
-\frac{\pi}{24 R}.
\label{a2.20}
\end{equation}
to the sum of the divergent series (\ref{a2.11}). In the theory
of divergent series~[15] this summation of the series
(\ref{a2.11}) is referred to as the
{\it Ramanujan summation}. Obviously, this method is not universal.
For
example, it cannot be applied directly to a divergent series
$\sum_{n=1}^{\infty}n^{-1}$ because  the zeta-function has a pole
at the point $s=1$ (see Eq.~(\ref{a2.15})). Admitting the convention
about the rejection of the pole singularity, as it is usually done in
the analytic regularization method\footnote{It is implicitly assumed
that the singularities are taken away by renormalization of
parameters in the theory  under consideration.}, we get
$$\sum\limits_{n=1}^{\infty}\frac{1}{n}=\gamma. $$
From (\ref{a2.15}), (\ref{a2.16}) it follows that the pole of the
Riemann $\zeta$-function at the point $s=1$ is responsible for the
logarithmic
divergences.

Summarizing, we arrive at the conclusion that
the Riemann $\zeta$-function method enables one to obtain the
finite value
of the Casimir energy (\ref{a2.11}) without explicit regularization,
pole singularity rejection, and explicit renormalization. However in
the case of divergent series (\ref{a2.12}) the analytic continuation
technique requires additional assumptions.

To sum the series (\ref{a2.12}), we have to consider the
Epstein-Hurwitz
zeta-function $\zeta_{EH}(s,p)$ defined by the formula
\begin{equation}
\zeta_{EH}(s,a^2)=\sum\limits_{n=1}^{\infty}(n^2+a^2)^{-s},
\label{a2.21}
\end{equation}
where $s>1/2$.
Let us remind briefly how to accomplish an analytic continuation of
this series to the region $s\le 1/2$.

Using the integral representation for the Euler gamma function [14]
\begin{equation}
(n^2+a^2)^{-s}\Gamma(s)=\int\limits_{0}^{\infty}t^{s-1}
e^{-(n^2+a^2)t}dt,
\label{a2.22}
\end{equation}
we can replace each term in  series (\ref{a2.21}) by the integral
\begin{equation}
\zeta_{EH}(s,a^2)=\frac{1}{\Gamma(s)}\int\limits_{0}^{\infty}t^{s-1}
\sum\limits_{n=1}^{\infty}e^{-t(n^2+a^2)}dt.
\label{a2.23}
\end{equation}
The Jacobi $\theta$-function appearing in (\ref{a2.23}) $\theta(t)=
\sum_{n=1}^{\infty}e^{-n^2 t}$ has the property~[14]
\begin{equation}
\theta(t)=-\frac{1}{2}+\frac{1}{2}\sqrt{\frac{\pi}{t}}+
\sqrt{\frac{\pi}{t}}\,\theta(\pi^2/t).
\label{a2.24}
\end{equation}
Substituting (\ref{a2.24}) into (\ref{a2.23}) we obtain
$$
\zeta_{EH}(s,a^2)=-\frac{(a^2)^{-s}}{2}+\frac{\sqrt{\pi}
\Gamma(s-1/2)}
{2\Gamma(s)}(a^2)^{-s+1/2}+$$
\begin{equation}+\frac{\sqrt{\pi}}{\Gamma(s)}\sum
\limits_{n=1}^{\infty}\int\limits_{0}^{\infty}t^{s-3/2}
\exp\left(-t a^2-\frac{\pi^2 n^2}{t}\right)dt.
\label{a2.25}
\end{equation}
The multiplier $\exp(-t a^2-\pi^2 n^2/t)$ ensures the convergence of
the integral in (\ref{a2.25}) for all $s$. Then this integral is
expressed in terms of the modified Bessel function
$K_{\nu}(z)$ having the integral representation~[16]
\begin{equation}
\int\limits_{0}^{\infty}x^{\nu-1}\exp\left(-\frac{\gamma}{x}-\delta x
\right)dx=
2\left(\frac{\gamma}{\delta}\right)^{\nu/2}K_{\nu}(2\sqrt{\gamma
\delta}),
\label{a2.26}
\end{equation}
$$K_{-\nu}(z)=K_{\nu}(z).$$
In the case under consideration, $\gamma$ and $\delta$ are positive
quantities:
$\gamma= \pi^2 n^2,~\delta=a^2.$
Now we can rewrite formula (\ref{a2.25}) as follows
$$
\zeta_{EH}(s,a^2)=-\frac{(a^2)^{-s}}{2}+\frac{\sqrt{\pi}\Gamma(s-1/2)}
{2\Gamma(s)}(a^2)^{-s+1/2}+$$
\begin{equation}
+\frac{2\pi^s}{\Gamma(s)}(a^2)^
{-s/2+1/4}
\sum\limits_{n=1}^{\infty}n^{s-1/2}K_{s-1/2}
(2\pi n \sqrt{a^2}).\label{a2.27}
\end{equation}
The series obtained converges for all $s$ as the modified Bessel
function has the  asymptotics~[16]
\begin{equation}
K_{\nu}(z)\sim\left(\frac{\pi}{2 z}\right)^{1/2}e^{-z},~~|z|\to\infty.
\label{a2.28}
\end{equation}
Therefore, the singularities of the function $\zeta_{EH}(s,a^2)$ are
due to the singularities of  $\Gamma(s-1/2)$ in (2.27), i.e.,
$\zeta_{EH}(s,a^2)$ has first order poles at the points
\begin{equation}
s=~\frac{1}{2},~ -\frac{1}{2},~ -\frac{3}{2},\cdots {\,}.
\label{a2.29}
\end{equation}
Thus formula~(\ref{a2.27}) affords an analytic continuation
of~(\ref{a2.21}) to the region  $s\le1/2$ except for the
points~(\ref{a2.29}).
Since $\zeta_{EH}(s,a^2)$ has a pole at $s=-1/2$,
function~(\ref{a2.27})
can be used for obtaining only the regularized Casimir energy
$E_C^{(2)\;reg}$
\begin{equation}
E_C^{(2)\;reg}=-\frac{M_0}{4\sqrt{\alpha}}-\frac{M_0^2 R}{8\pi\alpha}
\Gamma(-1)-\frac{M_0}{2\pi\sqrt{\alpha}}\sum\limits_{n=1}^{\infty}
n^{-1}
K_1\left(\frac{2n M_0 R}{\sqrt{\alpha}}\right).
\label{a2.30}
\end{equation}
In order for the regularization to be removed the first and  second
terms in the right-hand side of (\ref{a2.30}) should be omitted
(see, for example, [17--20])
\begin{equation}
E_C^{(2)\;ren}=-\frac{M_0}{2\pi\sqrt{\alpha}}\sum\limits_{n=1}^{
\infty}n^{-1}
K_1\left(\frac{2n M_0 R}{\sqrt{\alpha}}\right).
\label{a2.31}
\end{equation}
Rejection of the second term in (\ref{a2.30}) proportional to
$\Gamma(-1)$ is natural in the analytic continuation method\footnote{
If fields are considered in a bounded  space region, then this
procedure is interpreted as the subtraction of  infinite space
contribution~[21].}. As for the first term $-M_0/(4\sqrt{\alpha})$,
its rejection seems to be rather arbitrary.
Usually this is motivated by the fact that this term is independent
of $R$ and, as a consequence, does not contribute to the Casimir
force.
However, this argument does not explain the rejection of the
$R$-independent term in the interquark potential,
i.e., in $E_C^{(2)}(R).$ In the general case those terms may be
essential
for the description of quark-quark interaction inside  hadrons.
Only the consistent renormalization with preliminary regularization
and subsequent subtraction can justify the rejection of both the
first and  second terms in the right-hand side of (\ref{a2.30}).
This will be demonstrated in the next Section.
\section{Renormalization of the string tension and removal of the
divergences}
\setcounter{equation}{0}

Let us calculate the interquark potential (\ref{a2.5}), (\ref{a2.10})
applying the standard renormalization technique. The initial model
includes
two parameters: string tension $M_0^2$ and a dimensionless constant
$\alpha$ characterizing the string rigidity.  In the one-loop
approximation,
only the string tension is renormalized.

The renormalized potential  of the string at large distances should
coincide with
its classical  expression\footnote{In the framework of a string model,
the potential linearly rising at large distances is the classical
string energy considered as a function of its length $R$ when
$R\to\infty$~[6]. On the microscopic level (QCD level), the very
appearance of the collective  string degrees of freedom is
interpreted as a complicated nonperturbative effect in quantum
dynamics of gluon and quark fields closely related to nontrivial
properties of the QCD vacuum.}
\begin{equation}
V^{ren}(R)\left|_{R\to\infty}\right.=M^2 R,
\label{b3.1}
\end{equation}
where $M^2$ is the renormalized string tension, whose explicit
expression for  be obtained further. Starting with (\ref{a2.10})
and taking
into account the necessity to regularize all the divergent
expressions, we  represent $V^{ren}(R)$ as
\begin{eqnarray}
V^{ren}(R)&=&M_0^2 R+(D-2)\left .\left[E_C^{(1)\;reg}(R,\Lambda)+
E_C^{(2)\,reg}(R,\Lambda)\right ]\right |_{\Lambda\to\infty}=
\nonumber\\
&=& M_0^2 R+(D-2)
\left \{ \left[E_C^{(1)\;reg}(R,\Lambda)+
E_C^{(2)\;reg}(R,\Lambda)\right]\right.-\nonumber\\
&{}&-\left.\left.\left[E_C^{(1)\,reg}(R\to\infty,\Lambda)+
E_C^{(2)\,reg}(R\to\infty,\Lambda)\right ]
\right \}\right|_{\Lambda\to\infty}+\nonumber\\
&{}&+(D-2)
\left.\left[E_C^{(1)\;reg}(R\to\infty,\Lambda)+
E_C^{(2)\;reg}(R\to\infty,\Lambda)\right]\right|_{\Lambda\to\infty}=
\nonumber\\
&=&M^2 R+(D-2)\left[E_C^{(1)\;ren}(R)+
E_C^{(2)\;ren}(R)\right],\label{b3.2}
\end{eqnarray}
where $\Lambda$ is a regularization parameter;
$M^2$ is the renormalized value of the string tension,
\begin{equation}
M^2=M_0^2+\frac{D-2}{R}\left.\left[E_C^{(1)\;reg}(R\to\infty,\Lambda)
+E_C^{(2)\;reg}
(R\to\infty,\Lambda)\right]\right|_{\Lambda\to\infty};
\label{b3.3}
\end{equation}
and $E_C^{(i)\;ren},~i=1,2$ are the renormalized Casimir energies
(\ref{a2.11}) and (\ref{a2.12})
\begin{equation}
E_C^{(i)\;ren}(R)=\left.\left[E_C^{(i)\;reg}(R,\Lambda)-E_C^{(i)\;reg}
(R\to\infty,\Lambda)
\right]\right|_{\Lambda\to\infty},~~i=1,~2.
\label{b3.4}
\end{equation}

To regularize  divergent series (\ref{a2.11}) and (\ref{a2.12}),
we  substitute them by finite sums that can be represented
in terms of the Cauchy integrals~[14]
\begin{equation}
\frac{1}{2\pi i} \oint\limits_{C}^{}z\frac{f'(z)}{f(z)}dz
=\sum\limits_{k}^{}n_k a_k-\sum\limits_{l}^{}p_l b_l.
\label{b3.5}
\end{equation}
Here $f(z)$ is an analytic function having, in a region surrounded by
contour $C$, zeroes of order $n_k$ at points $z=a_k$ and poles of
order
$p_l$ at points $z=b_l$. As a function $f(z)$ we  substitute
 the right-hand sides of frequency equations
(\ref{a2.8}) and (\ref{a2.9})  into~(\ref{b3.5}) and choose the
contour $C$ so as
 to include $N$ first positive roots of the corresponding
equations.
Functions (\ref{a2.8}) and (\ref{a2.9}) have zeroes of the first
order on the real axis and have no poles. Therefore only the first sum
with $n_k=1$ remains in the right-hand side of (\ref{b3.5}).

First, we obtain  the regularized Casimir
energy~(\ref{a2.11})
\begin{equation}
E_C^{(1)\,reg}(R)=\frac{R}{4\pi i} \oint\limits_{C}^{} \omega\,
\frac{\cos{(\omega R)}}{\sin{(\omega R)}}\, d \omega,
\end{equation}
where the contour $C$ is shown in Fig.~2.   All the singularities of
the integrand in~(3.6) being situated on the real axis, it is possible
to deform the contour $C$ to $C'$ continuously (see Fig.~2)
Now the  regularization  parameter  is the radius $\Lambda$ of the
semicircle
entering into the contour $C'$.

To determine the counterterms according to (\ref{b3.1})--(\ref{b3.4}),
it is necessary to find the asymptotics of $E_C^{(1)\,reg}(R)$  for
$R\to\infty$ and fixed $\Lambda$. On the semicircle of radius
$\Lambda$ (Fig.~2) the asymptotics of the integrand for
$R\to\infty$ is the integrand itself because of its oscillating
character. Consequently, the result of integration along this part
of the
counter $C'$ is completely absorbed by the counterterm and does not
give any finite contribution to $E_C^{(1)ren}(R)$. Now let us turn
to the
integral along the interval $(-i\Lambda,i\Lambda)$ on the imaginary
axis
\begin{equation}
E_C^{(1)\;reg}(R,\Lambda)=-\frac{R}{4\pi}
\int\limits_{-\Lambda}^{\Lambda}
y\,\frac{\cosh(R y)}{\sinh(R y)}\,d y.
\label{b3.7}
\end{equation}
To find the asymptotics needed, we integrate in (3.7) by parts
$$
E_C^{(1)\;reg}(R,\Lambda)=-\frac{1}{4\pi}
\int\limits_{-\Lambda}^{\Lambda}y~
d\,(\ln|\sinh(R y)|)=
-\frac{\Lambda}{2 \pi}\ln[\sinh(\Lambda R)]+\frac{1}{2\pi}
\int\limits_0^{\Lambda}dy\ln[\sinh(Ry)].
$$
When $R\to\infty$
\begin{equation}
E_C^{(1)\;reg}(R\to\infty,\Lambda)=-\frac{\Lambda}{2\pi}
\ln[\sinh(\Lambda R)]+
\frac{1}{2 \pi}\int\limits_{0}^{\Lambda}(Ry-\ln 2)~dy.
\label{b3.9}
\end{equation}
Inserting (3.8) into (3.4) we obtain the finite value
for the renormalized Casimir energy (2.11)
\begin{equation}
E_C^{(1)\;ren}(R)=\frac{1}{2\pi}\int\limits_{0}^{\infty}
\ln\left(1-e^{-2R\omega}\right)~d\omega=-\frac{R}{\pi}\int
\limits_{0}^{\infty}
\frac{\omega~d\omega}{e^{2R\omega}-1}.
\label{b3.10}
\end{equation}
The last formula is derived by integrating by parts. It is interesting
to note that (3.9) is expressed in terms of the value of the
Riemann $\zeta$-function at the point $s=2$. Really,
\begin{equation}
\int\limits_{0}^{\infty}\frac{\omega d\omega}{e^{a\omega}-1}=
\int\limits_{0}^{\infty}\frac{\omega e^{-a\omega}}{1-e^{-a\omega}}~d
\omega=
\sum\limits_{n=1}^{\infty}\int\limits_{0}^{\infty}\omega
e^{-an\omega}d\omega=
\frac{\Gamma(2)}{a^2}\sum\limits_{n=1}^{\infty}\frac{1}{n^2}
=\frac{1}{a^2}\zeta(2).
\label{b3.11}
\end{equation}
In view of this, Eq.~(\ref{b3.9}) can be rewritten as
\begin{equation}
E_C^{(1)\;ren}(R)=-\frac{1}{4\pi R}\zeta(2).
\label{b3.12}
\end{equation}
Thus, under consistent renormalization, the sum of  divergent
series (\ref{a2.11}) is also defined through the Riemann $\zeta$-function,
but now another  range of its definition is used, namely, the region
$\mbox{Re}\,s>1$. Here $\zeta(s)$ is defined by convergent
series~(\ref{a2.13}).

With the help of the Riemann reflection formula (\ref{a2.17}) the value
of $\zeta$-function at $s=2$ entering into (3.11) can be
expressed through $\zeta(-1)$
\begin{equation}
\zeta(2)=-2\pi^2\zeta(-1).
\label{b3.13}
\end{equation}

Final renormalized formula for the Casimir energy~(\ref{a2.11}) assumes
the same form as that obtained by analytic continuation of
the Riemann $\zeta$-function ( Eq. (\ref{a2.20}))
\begin{equation}
E_C^{(1)\;ren}(R)=\frac{\pi}{2R}\zeta(-1)=\frac{\pi}{2R}
\left(-\frac{1}{12}\right)=-\frac{\pi}{24R}.
\label{b3.14}
\end{equation}

Thus, there is a complete agreement between two outlined  approaches
to the calculation of the  finite value of $E_C^{(1)}$.

Before we turn to consideration of the series (\ref{a2.12}),
let us make a short remark concerning formula (\ref{b3.10}).
Discarding the minus sign, the integrand in~(\ref{b3.10})
has a form of the Planck energy distribution  in the spectrum
of one-dimensional black-body with temperature $1/2 R$.

The renormalized value of the Casimir energy $E_C^{(2)}$
(see Eqs. (\ref{a2.12}), (\ref{b3.4})) can be obtained in the same way
as it was done above.
Substitution of the frequency equation~(\ref{a2.9}) into~(3.5)
gives
\begin{equation}
…_C^{(2)\;reg}(R,\Lambda)=\frac{R}{4\pi i}\oint\limits_{C}^{}
\frac{\cos(\sqrt{\omega^2-\omega_0^2})}
{\sin(\sqrt{\omega^2-\omega_0^2})}~
\frac{\omega^2}{\sqrt{\omega^2-\omega_0^2}}~d\omega,
\label{b3.15}
\end{equation}
where $\omega_0^2=M_0^2/\alpha$.
When choosing the contour $C$ one has to take into account the branch
points of the integrand $\omega=\pm\omega_0$. To select the
single-valued branch of the function, we make a cut connecting
the branch points along the real axis. After that the contour can be
chosen as shown in Fig.~3. Again integration along the semicircle of
radius~$\Lambda$  contributes only to the counterterm.
The integrals along the edges of the cut are mutually cancelled,
and the contribution $I_1$ of integration around the branch point
$\omega=\omega_0$ is equal to $-\omega_0/4=-M_0/(
4\sqrt{\alpha})$ (see Appendix~A).  It should be noted that
$I_1$ is exactly  equal  to the first term in
formula~(\ref{a2.30}) for $E_C^{(2)\;reg}$.
The sum of integral $I_2$ along the interval $(-i\Lambda,i\Lambda)$
of the imaginary
axis and $I_1$ is
\begin{equation}
E_C^{(2)\;reg}(R,\Lambda)=-\frac{\Lambda}{2\pi}\ln\left[
\sinh\left(R\sqrt{\Lambda^2+\omega_0^2}\right)\right]+
\frac{1}{2\pi}\int
\limits_{0}^{\Lambda}dy\ln\left[\sinh\left(R\sqrt{y^2+\omega_0^2}
\right)\right]-
\frac{\omega_0^2}{4}.
\label{b3.16}
\end{equation}
Integration by parts is already done here. Formula~(\ref{b3.4})
requires an asymptotics $E_C^{(2)\;reg}(R\to\infty,\Lambda)$.
From~(\ref{b3.16}) it follows that
\begin{equation}
E_C^{(2)\;reg}(R\to\infty,\Lambda)=-\frac{\Lambda}{2\pi}\ln\left[
\sinh\left(R\sqrt{\Lambda^2+\omega_0^2}\right)\right]+
\frac{1}{2\pi}\int
\limits_{0}^{\Lambda}dy\left(R\sqrt{y^2+\omega_0^2}-\ln 2\right)-
\frac{\omega_0^2}{4}.
\label{b3.17}
\end{equation}
The constant term $-\omega_0^2/4$ is preserved here to satisfy
condition~(\ref{b3.1}) which defines the behavior of the string
potential at infinity. Otherwise this term would appear in the
right-hand side of~(\ref{b3.1}), but that is physically unacceptable.
At large distances
string potential should be determined by its classical value only.
Inserting~(\ref{b3.16}) and (\ref{b3.17}) into (\ref{b3.4}) we find
for $i=2$
\begin{equation}
E_C^{(2)\;ren}(R)=\frac{1}{2\pi}\int\limits_0^{\infty}d\omega
\ln\left(1-
e^{-2R\sqrt{\omega^2+\omega_0^2}}\right)=
-\frac{R}{\pi}\int
\limits_0^{\infty}\frac{\omega^2 d\omega}{\sqrt{\omega^2+\omega_0^2}}~
\frac{1}{e^{2R\sqrt{\omega^2+\omega_0^2}}-1}.
\label{b3.18}
\end{equation}

It is interesting to the compare the formula derived with an analogous
expression for
$E^{(1)\;ren}_C(R)$ (see Eq. (3.9)). Formula (3.17) can be
obtained  by changing the variable frequency in~(3.9)
 to $\sqrt{\omega^2+\omega_0^2}$. This completely corresponds
to the fact that  Eq. (\ref{b3.10}) deals with oscillations of the
massless (two-dimensional) scalar field on the segment  $[0,R]$
while Eq.~(\ref{b3.18}) treats oscillations of the same field, but
with the mass equal to $\omega_0=M_0/\sqrt{\alpha}$
(see field equations (\ref{a2.6})).

At first sight, the expression obtained for $E_C^{(2)\;ren}(R)$ by
making
use of the consistent renormalization of the string tension does
not coincide with that derived by  analytic continuation of the
Epstein-Hurwitz zeta function (see formula (\ref{a2.31})).
This is not true, however. Equations~(\ref{b3.18}) and (\ref{a2.31})
are completely equivalent. To show this, let us expand the logarithm
in~(\ref{b3.18})
\begin{equation}
E_C^{(2)\;ren}(R)=-\frac{1}{2\pi}\sum\limits_{n=1}^{\infty}n^{-1}
\int\limits_{0}^{\infty}d\omega\, e^{-2nR\sqrt{\omega^2+\omega_0^2}}.
\label{b3.19}
\end{equation}
By changing the variable $\omega=\omega_0\sinh t$, the integral is
reduced
to  the tabular one~[16]
$$ K_{\nu}(z)=\int\limits_{0}^{\infty}e^{-z\cosh t}\cosh(\nu t)\, dt
$$
with $ z=2n\omega_0 R $. Finally we deduce  the series (\ref{a2.31})
from (\ref{b3.18})
\begin{equation}
E_C^{(2)\;ren}(R)=\frac{1}{2\pi}\int\limits_{0}^{\infty}d\omega\ln
\left(1-e^{-2R\sqrt{\omega^2+\omega_0^2}}\right)=
-\frac{\omega_0}{2\pi}
\sum\limits_{n=1}^{\infty}n^{-1}K_1(2n\omega_0 R),
\label{b3.20}
\end{equation}
where $\omega_0=M_0/\sqrt{\alpha}$. Thus we found an integral
representation for the series (\ref{a2.31}).
This series is convenient for investigating the behavior of
the Casimir energy $E_C^{(2)\;ren}(R)$ at large distances. Taking into
consideration~(\ref{a2.28}) we get
\begin{equation}
\left . E_C^{(2)\;ren}(R)\right|_{R\to\infty}\simeq-\frac{1}{4}\left(
\frac{\omega_0}{\pi R}\right)^{1/2} e^{-2\omega_0 R}.
\label{b3.21}
\end{equation}
The integral representation (3.17) enables one to study the
asymptotics of $E_C^{(2)\;ren}(R)$ at small $R$. From (\ref{b3.18})
it follows that $E_C^{(2)\;ren}(R)$ has a singularity when $R=0$.
For small $R$ the main contribution to this  integral
is given by large  $\omega$, therefore one can neglect here the
dependence
on $\omega_0$.
This immediately gives the asymptotics of  $E_C^{(2)ren}(R)$ for
$R\to0$
\begin{equation}
\left .E_C^{(2)\;ren}(R)\right |_{R\to0}\simeq\frac{1}{2\pi}
\int\limits_{0}^{\infty}
d\omega\ln\left(1-e^{2\omega R}\right)=-\frac{\pi}{24R}.
\label{b3.22}
\end{equation}

Thus, consistent regularization of the divergent series~(\ref{a2.12})
and subsequent renormalization of the string tension  justify the
rejection of
the singular (pole) term and $R$-independent constant in
Eq.~(\ref{a2.30})
when analytic continuation of Epstein-Hurvitz $\zeta$-function is
used.
It is worthwhile to emphasize an important advantage of the
proposed regularization by contour integration and subsequent
subtraction. In this way we obtain the spectral representation for
string energy at zero temperature (see Eqs.~(\ref{b3.10}) and
(\ref{b3.18}) in contrast to analytic continuation of
$\zeta$-functions
(Eqs.~(\ref{a2.20}) and (\ref{a2.31})). Proceeding from this spectral
representation one can immediately derive the string free energy at
finite temperature.
To this end one must pass from integration to summation
over the Matsubara frequencies
$\Omega_n=2\pi nT,~n=0,\pm1,\pm2,\dots$. Practically it is done by the
substitution
\begin{equation}
d\omega\to2\pi T\,d\omega\mathop{{\sum}'}\limits_{n=0}^{\infty}
\delta(\omega-\Omega_n),
\label{b3.23}
\end{equation}
where $T$ is the temperature (see Appendix~B).
The prime of the sum sign means that the term with $n=0$ should be
multiplied by $1/2$.

For either  quantity $E_C^{(i)}(R)\; i=1,2$ we have obtained two
integral
representations (see Eqs.~(\ref{b3.10}) and (\ref{b3.18})).
Substitution (3.22) in these
formulas with logarithmic functions gives us the {\it free
energy} at
finite temperature (Appendix~B). For example,
\begin{equation}
 F^{(2)}(R,T)=2T\mathop{{\sum}'}\limits_{n=0}^{\infty}\ln
\left(1-e^{-2R\sqrt{\Omega_n^2+\omega_0^2}}\right).
\end{equation}
Taking the limit $\omega_0\to0$ in (3.23) one can obtain the
free energy $F^{(1)}(R,T)$  that diverges due to the term with $n=0$
(see Appendix B). Making the substitution (3.22) in the second
version  of the spectral representations (3.9) and (3.17)
we arrive at the {\it internal energy} at temperature~$T$
\begin{equation}
U^{(1)}(R,T)=-4\pi R T^2\mathop{{\sum}'}\limits_{n=0}^{\infty}
\frac{n}{\exp(4\pi n R T)-1},
\label{b3.24}
\end{equation}
\begin{equation}
U^{(2)}(R,T)=-8\pi^2 R T^3\mathop{{\sum}'}\limits_{n=0}^{\infty}
\frac{n^2}{\sqrt{\Omega_n^2+\omega_0^2}}~
\frac{1}{\exp(2R\sqrt{\Omega_n^2+\omega_0^2}) -1}.
\label{b3.25}
\end{equation}
Both the energies, $U^{(i)}(R,T),\; i=1,2$ are well defined. The last
two equations prove to be convenient for investigating the behaviour
of the  internal energies at  large and small~$T$. Let us demonstrate
this
using Eq.~(3.24).
At large $T$ the main contribution to
(3.24) comes from the first term with $n=0$
\begin{equation}
U^{(1)}(R, T\to \infty)= - \frac{T}{2}\,{.}
\end{equation}
At small $T$ the Euler-Maclaurin  formula
\begin{equation}
\mathop{{\sum}'}_{n=0}^{\infty}f(n)= \int\limits_{0}^{\infty}f(x)
 \, dx -
\frac{1}{12} f'(0)
\end{equation}
can be used.
In the case under consideration
\begin{equation}
f(x)=\frac{x}{\exp{(4\pi T R x)}-1}\quad  \mbox{and}\quad f'(0)
=-\frac{1}{2}\,{.}
\end{equation}
As a result, we obtain for small~$T$
\begin{equation}
U^{(1)}(R,T)\approx  -\frac{\pi}{24\, R}- \frac{\pi T^2R}{6}\,{.}
\end{equation}

\section{Conclusion}
The experience of treating the divergences shows that a correct result
can be obtained by applying practically any regularization and
renormalization
procedures provided that the prescriptions are properly modified.
Therefore, when evaluating such
methods, those should be preferred which are closer to the quantum
field  theory.
Only in the framework of this approach one succeeds in formulation of
a consistent renormalization procedure. Besides, quantum field
formalism provides a clear and simple transition from zero temperature
calculations to those at finite temperature~[23].
In view of this, contour integration has an obvious advantage.
At first it was proposed as a simple method for calculating the
van der Waals forces between dielectrics~[24] (see also~[13, 17, 25,
26]). However its relation to the formalism of  the Green's functions is not
elucidated properly. And this problem is undoubtedly worth
investigating.
\section*{Acknowledgments}
The present work has been completed during the stay of one of the
authors (V.V.N.) at Salerno University.
It is a pleasant duty for him to thank Prof. G.~Scarpetta,
Dr.~G.~Lambiase and Dr.~A.~Feoli
for their kind hospitality and valuable discussions of the problems
touched upon in this paper.
The assistance of A.V.~Nesterenko in preparing
the figures is acknowledged  with gratitude.

\section*{Appendix A. Investigation of the contour integral
determining  $E_C^{(2)\;reg}(R,\Lambda)$}
\renewcommand{\theequation}%
{A.\arabic{equation}}
\setcounter{equation}{0}

Let us consider integral (3.14) for individual parts of contour
$C$ shown in  Fig.~3. Integration along the semicircle of  radius
$\Lambda$ contributes only to the counterterm, therefore we do
not analyze it here. When going from the upper edge of the
cut to the lower
one, the integrand does not change. As a result, integrals along these
two parts of the contour $C$ are cancelled mutually due to  opposite
directions of integration.
Only  integration around the branch point and along the interval of
the imaginary axis $(-i\Lambda,i\Lambda)$ lead to finite
contributions. While integrating around the branch point
$\omega=\omega_0$ we introduce
usual variables $\omega-\omega_0=\rho e^{i\phi}$
with~{$\rho\to 0$}, and in terms of them we have
$\omega^2-\omega_0^2=(\omega+
\omega_0)(\omega-\omega_0)\simeq 2 \omega_0 \rho e^{i\phi},\;\;
\cos(R\sqrt{\omega^2-\omega_0^2})\simeq 1,\;\;
\sin(R\sqrt{\omega^2-\omega_0^2})
\simeq R\sqrt{\omega^2-\omega_0^2}$.
Taking this into account we deduce
\begin{equation}
I_1=-\frac{R}{4\pi i}\int\limits_{0}^{2\pi}\frac{\omega_0^2 \rho
e^{i\phi} i\, d\phi}{2\omega_0 \rho e^{i\phi}R}=-\frac{\omega_0}{4}
=-\frac{M_0}{4\sqrt{\alpha}}.
\end{equation}
The integral $I_1$ is exactly equal to the first term in
(\ref{a2.30})  which is  independent of  $R$.
When integrating along the imaginary axis,  trigonometrical
functions in (3.14) become hyperbolic ones
\begin{equation}
I_2=\frac{R}{4\pi i}\int\limits_{\Lambda}^{-\Lambda}
\frac{(-y^2)\,i\, dy}{i\sqrt{y^2+\omega_0^2}}\frac{\cosh
(R\sqrt{y^2+\omega^2_0})}{\sinh
(R\sqrt{y^2+\omega^2_0})}dy=
-\frac{1}{2\pi}\int\limits_{0}^
{\Lambda}y\, d\left[\ln\sinh\left(R\sqrt{y^2+\omega_0^2}\right)
\right].
\end{equation}
Summing (A.1) and (A.2) and integrating by parts one arrives
at formula (3.15)
\section*{Appendix B. Transition to finite temperature in infinite
system of noninteracting oscillators}
\renewcommand{\theequation}%
{B.\arabic{equation}}
\setcounter{equation}{0}
Let us consider an infinite system of noninteracting oscillators with
eigenfrequencies $\omega_n,\;n=1, 2,\dots$ determined by the equation
\begin{equation}
f(\omega,R)=0.
\label{App1}
\end{equation}
 Roots of this
equation are assumed to be situated  on the real axis in
the complex plane $\omega$.
This set of oscillators arises, for example, in quantization of a
scalar field
defined on the line segment $[0,\,R]$. Boundary conditions
imposed on this field result in frequency equation~(B.1).
Without loss of generality, for relatistically invariant system one can
admit that the function $f$ satisfies the condition
\begin{equation}
f(-\omega,R)=f(\omega,R).
\label{App2}
\end{equation}

The free energy of this system is given  by
\begin{equation}
 F(R,T)=\sum \limits_{n=1}^{\infty}\left[\frac{\omega_n}{2}
+T\ln\left(1-e^{-\omega_n/T}\right)\right].
\label{App3}
\end{equation}
At zero temperature this formula obviously turns into the energy
of zero point oscillations $E_C=\sum_{n=1}^{\infty}\omega_n/2$.

In the general case  sum (\ref{App3}) diverges, therefore to
obtain the finite value for
the free energy, we have to use the renormalization
procedure discussed in Section 3. First, the infinite sum should be
represented as the contour integral
\begin{equation}
\sum\limits_{n=1}^{\infty}g(\omega_n)=\frac{1}{2\pi i}\oint\limits_C^{}
g(z)\frac{f'(z,R)}{f(z,R)}\,dz,
\label{App4}
\end{equation}
where
$$
g(\omega)=\frac{\omega}{2}+T\ln \left(1-e^{-\omega/T}\right)=
\frac{\omega}{T}-T\sum\limits_{n=1}^{\infty}\frac{1}{n}
e^{- n \omega/T}.
$$
The contour $C$, as in  Section 3, surrounds the first
$N$ roots of  Eq.~(\ref{App1}).

As shown  in Section~3, only integration along the imaginary axis gives
a finite contribution to the free energy
\begin{equation}
F^{reg}(R,\Lambda)=-\frac{1}{2\pi i}\int\limits_{-\Lambda}^{\Lambda}
\left(\frac{i y}{2}-T\sum\limits_{n=1}^{\infty}
\frac{1}{n}\,e^{-i n y/T}\right)\,d\ln[f(iy,R)].
\label{App5}
\end{equation}
On integrating by parts we obtain
\begin{equation}
F^{reg}(R,\Lambda)=\frac{1}{2\pi }\int\limits_{-\Lambda}^{\Lambda}
dy\left(\frac{1}{2}+\sum\limits_{n=1}^{\infty}e^{-i n y/T}\right)\,
\ln[f(iy,R)]=\label{App6}\end{equation}
$$=\frac{1}{2\pi }\int\limits_{-\Lambda}^{\Lambda}dy
\left\{\frac{1}{2}+\sum\limits_{n=1}^{\infty}\left[\cos\left(
\frac{n y}{T}
\right)-i\sin\left (\frac{n y}{T}\right )
\right]\right\}\,
\ln[f(iy,R)]$$
The off-integral terms are omitted in (\ref{App6}) because they
contribute only to the counterterm.
Taking into account (\ref{App2}) we can drop terms with sine functions
 \begin{equation}
 F^{reg}(R,\Lambda)=\frac{1}{\pi }\int\limits_{0}^{\Lambda}
dy \left[\frac{1}{2}+\sum\limits_{n=1}^{\infty}
\cos\left(\frac{n y}{T}\right)\right]\,
\ln[f(iy,R)]
\label{App7}
\end{equation}
The renormalized free energy is obtained by the subtraction
\begin{eqnarray}
 F^{ren}(R,T)&
=&\left .\left[ F^{reg}(R,\Lambda)-
F^{reg}(R\to\infty,\Lambda)\right]\right |_{\Lambda\to\infty}=
\nonumber\\
&=&\frac{1}{\pi }\int\limits_{0}^{\infty
}
dy \left[\frac{1}{2}+\sum\limits_{n=1}^{\infty}
\cos\left(\frac{n y}{T}\right)\right]\,
\ln\left[\frac{f(iy,R)}{f(iy,\infty)}\right].
\label{App8}
\end{eqnarray}
With allowance for  the Fourier-series representation
of the $\delta$-function
$$
\pi T\sum\limits_{n=-\infty}^{\infty}
\delta(y-2\pi n T)=
\frac{1}{2}+\sum\limits_{n=1}^{\infty}\cos\left(\frac{ny}{T}\right)
$$
integration in (\ref{App8})  can be done to produce
\begin{equation}
F^{ren}(R,T)=
T\sum\limits_{n=-\infty}^{\infty}
\ln\left[\frac{f(2\pi i n T, R)}
{f(2\pi i n T,\infty)}\right]=
2T\mathop{{\sum}'}\limits_{n=0}^{\infty}
\ln\left[\frac{f(2\pi i n T, R)}
{f(2\pi i n T,\infty)}\right].
\label{App9}
\end{equation}

Now we apply formula (\ref{App9}) to the models considered in
Section~3. In the case of a scalar field with mass $\omega_0$ on
the segment $[0,R]$ we have
$$ f(\omega,R)=\sin\left(R\sqrt{\omega^2+\omega_0^2}\right),
$$
and  Eq.~(\ref{App9}) gives
\begin{equation}
F^{(2)}(R,T)=2T\mathop{{\sum}'}\limits_{n=0}^{\infty}\ln
\left(1-e^{-2R\sqrt{\Omega_n^2+\omega_0^2}}\right),
\label{App10}
\end{equation}
where $\Omega_n=2\pi n T$.
The same result was obtained in Section~3 by   transition
from the integral representation for the Casimir energy  at zero
temperature
to summation over the Matsubara frequencies (see Eq.\ (3.23)).
Proceeding from (\ref{App10}) one can derive the $internal\; energy$
of  the system under consideration applying thermodynamic rules
\begin{equation}
U^{(2)}(R,T)=-T^2\left[\frac{\partial}{\partial T}
\frac{F^{(2)}(R,T)}{T}\right]=
-8\pi^2 R T^3\mathop{{\sum}'}\limits_{n=0}^{\infty}
\frac{n^2}{\sqrt{\Omega_n^2+\omega_0^2}}~
\frac{1}{\exp(2R\sqrt{\Omega_n^2+\omega_0^2}) -1}.
\label{App11}
\end{equation}
This equation was derived in Section~3 by a simple
substitution (see Eq. (3.25)).

In the case of massless scalar field $(\omega_0\to0)$ the term with
$n=0$ in (\ref{App10}) diverges
$$
F^{(1)}(R,T)=2T\mathop{{\sum}'}\limits_{n=0}^{\infty}\ln
\left(1-e^{-4\pi n R T}\right) =$$
$$=- T\lim_{n\to 0}\sum_{k=1}^{\infty}\frac{\exp(-4 \pi n k RT)}{k}+2T
\sum\limits_{n=1}^{\infty}\ln
\left(1-e^{-4\pi n R T}\right) = $$
\begin{equation}
=- T\sum_{k=1}^{\infty}\frac{1}{k} +
2T\sum\limits_{n=1}^{\infty}\ln
\left(1-e^{-4\pi n R T}\right)\,{.}
\label{App12}
\end{equation}
This divergence  is a manifestation of the well-known infrared
instability of a massless
scalar field in two-dimensional space-time.
In Section~2 some reasons were given to attribute
the Euler constant value, $\gamma$ to the sum of
the divergent series
$\sum_{k=1}^{\infty }k^{-1}$ .
Finally we obtain a finite expression for the free energy of the
massless scalar field
on the segment~$[0,\,R]$
\begin{equation}
F^{(1)}(R,T)= -\gamma T +2 T\sum _{n=1}^{\infty}\ln \left(1- e^{- 4
\pi n RT}\right )\,{.}
\end{equation}
It should be noted that this  treatment of infrared divergences
in the problem
in question  is absolutely formal and it needs the physical
justification.

However  the internal energy of this
 field  is well defined. Putting $\omega _0 =0$ in (B.11) we get
\begin{equation}
U^{(1)}(R,T)=-4\pi R T^2\mathop{{\sum}'}\limits_{n=0}^{\infty}
\frac{n}{\exp(4\pi n R T)-1}.
\label{App13}
\end{equation}
In Section~3 the same formula has been  derived  by a formal
substitution (see Eq.~(3.24)).

\newpage
\centerline{\large \bf Figure Captions}
\bigskip
\begin{description}
\item Fig.~1. Contour $C$ used in analytic continuation of the Riemann
zeta-function.
\item Fig.~2. Transformation  of the contour in integral~(3.6).
\item Fig.~3. Contour used for summing the roots of Eq.~(2.9).
\end{description}

\begin{thebibliography}{99}
\bibitem{fn1}Most
commonly this operator is the Laplace operator $(-\triangle)$. The
zeta function regularization is usually applied to the Euclidean version
of the models where one has to do with elliptic operators~[3].
\bibitem{fn2}{It is implicitly assumed
that the singularities are taken away by renormalization of
parameters in the theory  under consideration.}
\bibitem{fn3}If fields are considered in a bounded  space region, then this
procedure is interpreted as the subtraction of  infinite space
contribution~[24].
\bibitem{fn4}{In the framework of a string model, the
potential linearly rising at large distances is the classical
string energy considered as a function of its length $R$ when
$R\to\infty$ (see Ref.~7). On the microscopic level (QCD level), the very
appearance of the collective  string degrees of freedom is
interpreted as a complicated nonperturbative effect in quantum
dynamics of gluon and quark fields closely related to nontrivial
properties of the QCD vacuum.}
\end{thebibliography}
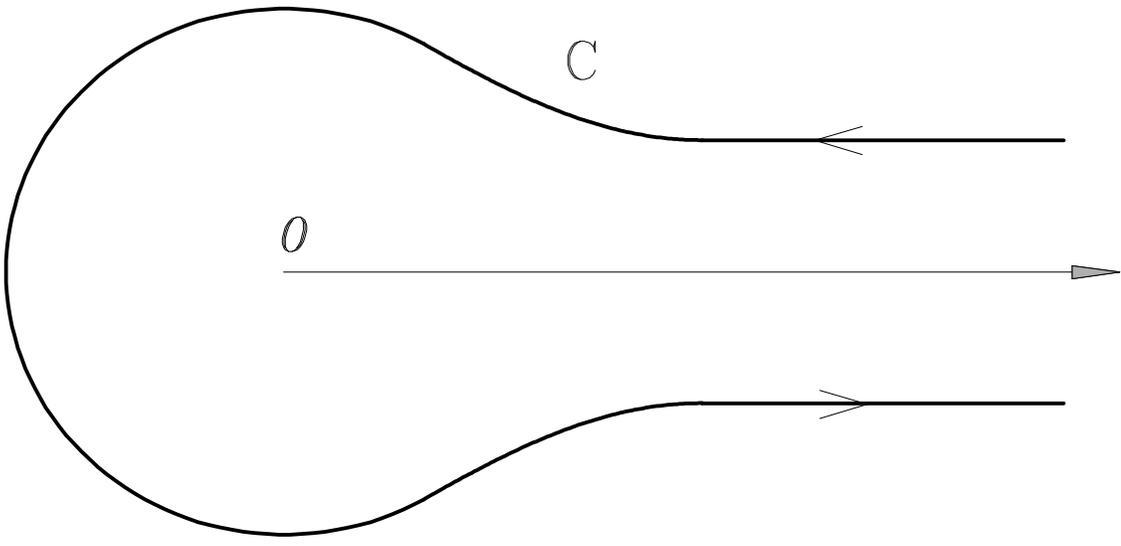
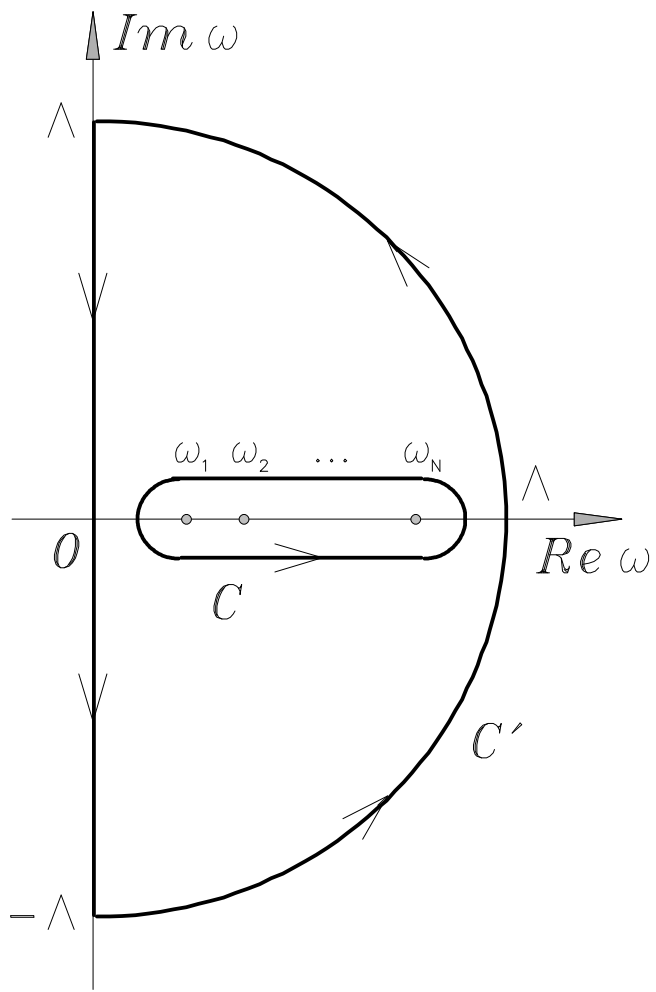
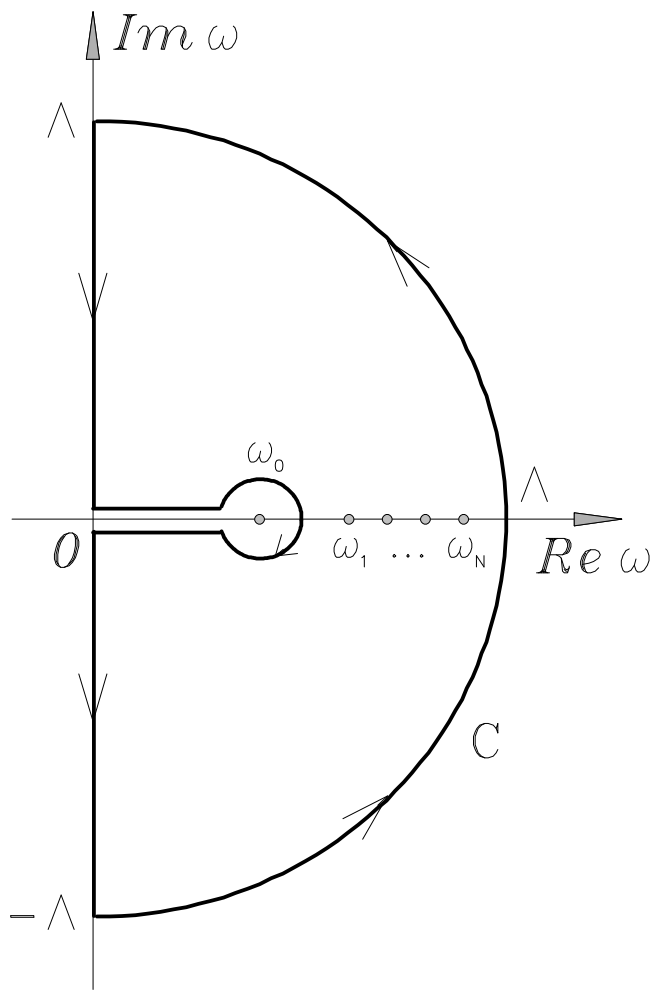
\end{document}